\documentclass[12pt,preprint]{aastex}

\shorttitle{Supernova Search in Globular Clusters}
\shortauthors{Washabaugh and Bregman}

\begin{document}

\title{The Production Rate of SN Ia Events in Globular Clusters}

\author{Pearce C. Washabaugh and Joel N. Bregman}
\affil{University of Michigan, Department of Astronomy, Ann Arbor, MI 48109}
\email{pearce@umich.edu and jbregman@umich.edu}

\begin{abstract}
In globular clusters, dynamical evolution produces luminous X-ray emitting binaries 
at a rate about 200 times greater than in the field.  If globular clusters also
produce SNe Ia at a high rate, it would account for much of the SN Ia events in early type
galaxies and provide insight into their formation.  
Here we use archival HST images of nearby galaxies that have hosted SNe Ia to examine 
the rate at which globular clusters produce these events. 
The location of the SN Ia is registered on an HST image obtained before the event or after the supernova faded.   
Of the 36 nearby galaxies examined, 21 had sufficiently good data to search for globular cluster hosts.
None of the 21 supernovae have a definite globular cluster counterpart, although there are some ambiguous cases.
This places an upper limit to the enhancement rate of SN Ia production in globular clusters of about 42 at the 95\% confidence level, which is an order of magnitude lower than the enhancement rate for luminous X-ray binaries. Even if all of the ambiguous cases are considered as having a globular cluster counterpart, the upper bound for the enhancement rate is 82 at the 95\% confidence level, excluding an enhancement rate of 200.
Barring unforeseen selection effects, we conclude that globular clusters are not 
responsible for producing a significant fraction of the SN Ia events in early-type galaxies.

\end{abstract}


\section{Introduction}

Not only are SN Ia events of interest as endpoints of stellar evolution, they 
have taken on a special importance because of their central role as standard candles.  
It was with these objects that the acceleration of the 
universe was first detected and they will likely play a central role in accurately 
measuring the acceleration parameters over cosmological time (e.g., \citealt{riess07, suzu12}).  This greater accuracy 
requires not only a large number of SNe Ia, but a good understanding of the supernova events.  
An important part of the picture is the set of conditions that cause white dwarfs 
to undergo a SN Ia event.

Within galaxies, close binary systems are rare and they are generally not created through dynamical interaction but are formed with a sufficiently small separation.  
However, a different situation occurs in globular clusters because 
the dynamical interaction time is typically 10$^{{\rm 8}}$-10$^{{\rm 9}}$ yr.  
These interactions cause binaries to harden to the point that the constituent stars become 
close enough for mass transfer (e.g., \citealt{hut92, pool03}).  
When this occurs, the star that expands beyond its Roche lobe experiences Roche 
lobe overflow and subsequent mass transfer to the more compact object.  When the compact 
object is a neutron star, we obtain a low mass X-ray binary (LMXB), which can 
be quite luminous in X-rays, making them easy to identify.  
Luminous X-ray binaries are seen in early-type galaxies, and many occur in 
globular clusters \citep{ang01,saraz03,irwin05,kundu07,siva07}.

The rate enhancement of LMXBs in globular clusters may be a critical quantity 
when estimating the number of Ia SNe that might be anticipated in these systems \citep{ivan06, pfahl09}.
In the Milky Way, about 1/10 of luminous X-ray binaries are in globular clusters
(typical X-ray luminosities exceeding $10^{36}$ erg sec$^{-1}$; e.g., \citealt{verb06}), 
and if one sums the globular clusters by mass \citep{mand91}
and compares it to the stellar mass for the Milky Way \citep{binney1998}, one finds that LMXBs
occur about 200 times more commonly than in an equivalent mass of field stars (the mass fraction
in globular clusters is 0.05\% based on these resources).
A similar enhancement can be determined for early-type galaxies, 
which are the galaxies considered in our sample.  
There is a difference compared to the Milky Way, because at the distances of
even nearby early-type galaxies, the binaries must be significantly more 
luminous in order to be detectable.  
These LMXBs have X-ray luminosities usually exceeding 10$^{37.5} erg sec^{-1}$ \citep{irwin05}.  
In these galaxies, about half of the X-ray binaries are in globular clusters \citep{saraz03, siva07}.  
Globular clusters are typically more common in early-type galaxies, relative to the Milky Way, 
which can be expressed by the specific frequency of globular clusters, $S_N$ \citep{harris91}.
The specific frequency is 0.5 for the Milky way, but is 2.6 for E/S0 in small groups and 5.4
for E/S0 galaxies in Virgo and Fornax \citep{harris91}; most of the galaxies in the Sivakoff sample
were in the Virgo cluster with the specific frequency in the range 3.5-14.
Using the data given in the Sivakoff sample, for a median $S_N$ = 5, an order of magnitude 
greater than the Milky Way, the stellar mass fraction in globular cluster stars is about 0.5\%. 
For about half of the LMXBs to be in globular clusters, this implies an enhancement rate 
of about 200 for sources more luminous than $2 \times 10^{38}$ erg sec$^{-1}$ in the 0.3-10 keV band.
We conclude that the enhancement rate of producing a LMXB in a globular cluster is about 200,
for X-ray luminosities in excess of $10^{36}$ erg sec$^{-1}$

If the fraction of SNe Ia that are hosted by globular clusters was the same as the 
fraction of LMXBs hosted by globular clusters, we would expect to find about 20-50\%
of SN Ia events in globular clusters, which is the proposition that we test here.
The differences between LMXBs and SN Ia progenitors are significant in that SN Ia 
events are associated with binaries involving a white dwarf (or two white dwarfs), 
rather than a neutron star (or black hole) plus a non-degenerate star, so the enhancement 
rate may be very different.
However, the frequency of white dwarf mass-transfer binaries also appears to be enhanced 
in globular clusters, based on X-ray observations \citep{heink03,pool06}.  
They are fainter than neutron star binaries (due to the difference 
in the depth of the potential well), but deep observations of nearby Galactic 
globular clusters indicates a large number of hard white dwarf binaries 
(Cataclysmic Variables).  Since globular clusters appear to be factories for 
the production of mass-transfer binaries, it is only natural to wonder if they 
might produce a significant fraction of the SNe Ia seen in early-type galaxies.  
In this paper we investigate the frequency with which such nearby SNe Ia are 
associated with globular clusters.  We accomplish this by examining the location
of these supernovae on \textit{Hubble Space Telescope} (HST) images that are 
sufficiently deep to detect globular cluster in external galaxies. 
In the next section we discuss the criteria by which we chose our targets. 
The following two sections give a summary of the results and notes on individual objects. 
The final section is a discussion of the results and implications for related efforts.
Recently, but after our initial presentation of results \citep{wash09}, \citet{voss12} 
published a similar effort, which we discuss in more detail in the final section.

\section{Target Selection}

In the selection of targets, it is crucial to work with systems that have 
well-defined positions for the SN. Early-type galaxies, the host sample, often have thousands of 
globular clusters (GCs) so the chance that the SN position will overlap with a 
random GC (or background source) unassociated with the SN becomes significant 
if the positional uncertainty is $r > 1\arcsec$ (positional uncertainties were 
taken from the NASA Extragalactic Database). 
If there are GCs in the possible region that the SN went off and the error circle is large because of a high positional uncertainty, then the event can at best be considered ambiguous because of the chance of random superposition.
Also, the region must have at least two guide stars
whose positions are well known so that the HST image can be accurately aligned.

Another criteria is that the system has to be close enough that one can detect most 
of the stars that lie in GCs.  This does not mean that all globular clusters be detected,
as the less luminous clusters add little to the total number of globular cluster stars.
For a luminosity function that is log normal, as found for the Milky Way and M31, the 
center of the Gaussian is at $M_V$ = $-$7.5 and with a dispersion of $\sigma$ = 1.25 
(e.g., \citealt{binney1998}).  For these parameters, 73\% of all globular cluster stars 
are in clusters brighter than $M_V$ =$ -$8 and 93\% in clusters brighter than $M_V$ =$-$7.
So, if they are brighter than about V = 25 mag, HST can obtain good photometric measurements 
in modest observing times. This requirement implies a distance modulus of 32.5 or less, or a distance 
limit of 32 Mpc, which we used as a selection criteria. 
Ground based images do not have the ability to reach the resolution and magnitude 
required for galaxies of this distance, making them inadequate for this project. 
We note that with sufficient S/N, the extended nature of most GCs will be detected at 
HST resolution \citep{jord05}, subtending $\sim 0.1-0.2\arcsec$ in radius.

A third consideration is that the search region of the host galaxy must be suitable for the detection of GCs. 
A galaxy with a very high surface brightness will compromise the ability to detect GCs 
and to determine that they are extended. In practical terms, one needs to avoid regions 
where the galaxy surface brightness exceeds 17 magnitudes per square arcsecond, 
otherwise the S/N of a typical globular cluster ($M_V = -7.5$) in a typical exposure falls below the detection
threshold (discussed further in section 4). 
Also, the density of GCs increases close to the core approximately like a de Vaucouleurs law, 
so a search region near the core has a much higher chance of superposition and such
matches are given as ambiguous detections.
For nearby early-type galaxies, these restrictions mean that there is significant 
incompleteness for Ia SNe within about 10-20$\arcsec$ of the galaxy center.  
Unfortunately, this requirement is often at odds with archival HST programs, which 
generally target the central part of galaxies, although there are a few directed 
programs to detect particular supernovae.
There were some HST observations to follow the light curve of particular SNe, and  
when there were also pre-SNe images (or 3+ yr post images), we could identify the 
position very accurately and these objects have the smallest error 
circles, typically 0.3$\arcsec$ or less (see below).

Finally, enough time needs to have passed since the SN Ia event so that it is 
fainter than the globular cluster we are trying to study. 
The peak absolute magnitude of a SN Ia is about $-$18 (in either B, V, or I) and it has to become 
fainter than $-$6, fading by at least 12 magnitudes. Based on known light 
curves (e.g., \citet{ries99,salv01,fran02}) 
and extrapolation from the known dimming due to radioactive half lives 
(or using the SN 1987a data), a dimming of 12 magnitudes occurs in a little less 
than 3 years in the optical bands. 

To sum up, the important factors governing the selection of targets are:
\begin{itemize}
\renewcommand{\labelitemi}{$\bullet$}
\item Positional uncertainty ( $>$1 arc sec may be ambiguous)
\item At least two guide stars in image
\item Galaxy is within 32 Mpc
\item SN Ia occurred within suitable location in galaxy: away from high surface brightness regions
\item The images must have been taken before or at least three years after the SN event occurred.
\item The host is an early-type galaxy
\end{itemize}

With these criteria, we searched all known historical Ia SNe through 2007 as listed
in the IAU Central Bureau for Astronomical Telegrams,\\ 
http://www.cbat.eps.harvard.edu/lists/Supernovae.html.
This resulted in the 36 Ia SNe events listed in Tables 1 and 2.

\section{Observational Results}

Of the 36 galaxies that satisfied our target criterion, only 21 had met our 
observational criteria, including:  the SN Ia fell within the HST footprint 
and not in a CCD gap (this is the primary criteria for excluding 15 events);
the observation was taken either before the supernova event at least three years 
after the event; astrometric objects are present on the image; and the supernova
position had adequate positional accuracy (Tables 1, 2).
Aside from one STIS image, where SN 1996X in NGC 5061 was observed in the 
50CCD mode (MIRVIS spectral element), the rest of the images were either 
from the ACS or the WFPC2 (Tables 3, 5, 6).  
For the WFPC2 and STIS materials, we used the data products from the \textit{Hubble Legacy Archive},
while for the ACS data, we used the standard drizzled image that is one of the data products,
available through MAST. See Table 3 for exposure times and filters.

To determine whether a SN Ia event corresponded to a globular cluster, we needed to align the
position of the SN with the HST image by translating and rotating the image so that 
at least two guide stars matched with positions given in the 2MASS survey.  
There can be an inaccuracy in the absolute position of an
HST image that can lead to improper identifications, so we pursued a few avenues to obtain the
best registration of the SN position on the HST image.
In a few cases, there was an HST image in which the SN was visible and another one, usually
taken before the SN occurred.  In these cases, an alignment to 0.1-0.3$\arcsec$ was achieved,
which are our most accurate results (e.g., SN 2003hv in NGC 1201).  
When two HST images were not available, we had to rely on the accuracy of the 
optical discovery images, which are often quoted to an accuracy of less than 1$\arcsec$.
In the discovery images, the positions for some of the SNe are measured 
relative to the center of the galaxy or to background point sources with known positions.
We checked the accuracy of the galaxy center being used, and when HST images included 
the galaxy center, we could use this information to position the error circle location.
More often, we performed astrometry corrections to the HST image for the alignment with the
optical SN position.  For the astrometry, we used the locations of stars in the field that 
occur in the 2MASS and the USNO-B1 catalogue, and the galaxy center, when available.
The size of the resulting error circle is the incoherent sum of the errors from the 
SNe discovery position plus the astrometry uncertainty. See Table 4 for error circle radii.

A source is a non-detection if no point source is found within or on the error circle.
A clear detection would be the result of a single object, detected at above 5$\sigma$, lying  
within the error circle, with a low probability that this occurs by chance from contaminating sources.
An ambiguous result refers to when either (1) the source within the error circle is 
in the 3-5$\sigma\ $ range, (2) there are multiple objects within the
error circle, (3) or when the density of sources is sufficiently high that a object can easily occur
within the error circle by chance.
With these criteria, in the sample of 21 objects, there were 16 clear non-detections and 
5 that were ambiguous. 
A more detailed object-by-object discussion is given below. 
The limiting magnitudes (Table 4) are often dictated by the brightness of the stellar background, which
was measured directly from the image local to the SN position.
We calculated the limiting magnitudes by using these backgrounds, along with the exposure time,
with the exposure time calculators provided at the HST site.
The median limiting magnitude was 24.4 mag, with a median distance modulus of 31.6.

\subsection{Object-By-Object Discussion}

We describe each event in our sample, listed chronologically.

SN 1919A, NGC 4486:  The SN positional uncertainty is 5.0$\arcsec$ since the SN occurred in 1919. This uncertainty is too large for our purposes.

SN 1960R, NGC 4382:  None of the WFPC2 and ACS images contain the SN position.

SN 1961H, NGC 4564:  The supernova occurs only 6$\arcsec$ from the galaxy center, so the 
galaxy surface brightness is high as is the number of globular clusters; 
chance coincidences are likely.
There is a dim object in the error circle (radius=1.7$\arcsec$) in the ACS images. 
Due to the large number of sources, we consider this association to be ambiguous.
The limiting magnitude is 23.0 in the F850LP band. See fig. 3

SN 1976K, NGC 3226:  A very precise ground-based position was provided by \cite{klemola1986},
with an uncertainty of only $\sim$0.1$\arcsec$. 
Using the ACS images, no object was found within the SN error circle (radius=0.4$\arcsec$) at our limiting magnitude of $25.1$ in the F814W band.

SN 1980i, NGC 4374:  None of the WFPC2 and ACS images contain the SN position.

SN 1982W, NGC 5485:  None of the WFPC2 and ACS images contain the SN position.

SN 1983G, NGC 4753:  None of the WFPC2 and ACS images contain the SN position.

SN 1983O, NGC 4220:  The region where the SN occurred in lies in the galactic disk and has a relatively large number of globular clusters. The SN has a positional uncertainty ($\approx$1$\arcsec$), thus our error circle (1.8$\arcsec$) encompassed several objects.
This constitutes a high level of contamination, making it impossible to associate the SN position
with a globular cluster. See fig. 2

SN 1986G, NGC 5128:  The SN is off of the only ACS image, but lies on three WFPC2 images. 
There are at least four objects in the SN error circle $(radius = 1.7\arcsec)$ and the region is in an area of heavy dust extinction. The chance of super-position with unrelated
sources is extremely high.  The uncertainty of the SN position is 
high enough that it is impossible to identify the SN with an underlying optical counterpart. See fig. 4

SN 1991bg, NGC 4374:  Three point sources in the WFPC2 and ACS images were used for astrometry and led to a good absolute positions for the HST images, but the uncertainty in the discovery images dominates the size of the error circle. 
There were no objects in the error circle (radius = 1.7$\arcsec$) in any images at our limiting magnitude of $24.4$.

SN 1991F, NGC 3485:  Using the SN position given by \citep{gomez95}, we determined that no WFPC2 or ACS images show a source at the location of the supernova to our limiting magnitude of $24.4$ in the F606W band.

SN 1992A, NGC 1380:  The ACS images was taken $12 - 14$ years after the SN and the astrometric sources aligned well with their nominal positions, so a positional shift was is unnecessary. This SN has a higher than average positional uncertainty ($1.5\arcsec$), leading to an error circle a radius of $2.5\arcsec$. 
Despite this relatively large error circle, no optical counterpart was found within the circle to our limiting magnitude of $24.0$ in the F555W band.

SN 1994D, NGC 4526:  Since the SN position is known to a high degree of accuracy (0.1$\arcsec$) we were able 
to constrain the error circle to only 0.4$\arcsec$. 
No objects were found in the error circle at the limiting magnitude of $23.2$ in the F850LP band.

SN 1995D, NGC 2962:  No WFPC2 or ACS images contain the SN position.

SN 1996X, NGC 5061:  The SN is visible in every WFPC2 image and appears very faint by 2000, the most recent WFPC2 image of the region. There is a STIS image of the region taken in 2002. We were able to align this image with the WFPC2 images so that the error circle has radius $0.2\arcsec$. The STIS image showed that there was no point source at the SN location to our limiting magnitude of 26.0 (the 50CCD setting was used).

SN 1996bk, NGC 5308: This is a flattened early-type galaxy in which the supernova lies in the midplane, about 20$\arcsec$ from the center and the surface brightness is moderately high.
There are several objects within the error circle on the ACS image, and there would be multiple objects within the $2.0\arcsec$ error circle if it were place almost anywhere along the disk of the galaxy.
Thus, there is severe contamination, making it impossible to determine if an optical object is
associated with the supernova. See fig. 5

SN 1997X, NGC 4691:  No HST images of this object exist.

SN 1999gh, NGC 2986:  There are two WFPC2 images of this object. One shows a potential cluster at the location of the SN, but the other does not confirm this result, implying that this source is a cosmic ray. Our error circle has a radius of $0.9\arcsec$. The limiting magnitude is $25.3$ and the filter was F702W.

SN 2000cx, NGC 524:  All WFPC2 images contain the SN when it was visible. The only ACS image, taken with the HRC, does not include the location of the SN. Thus there is no HST image of the region taken after the supernova has faded.

SN 2000ds, NGC 2768:  No object appears within our error circle (radius = 2.0$\arcsec$) to a limiting magnitude of $24.2$ with the F658N.

SN 2001A, NGC 4261:  The WFPC2 and ACS images were offset from each other by a few tenths of an arcsecond. 
Using the galactic center for astrometry and relative centering of the images, we found no optical counterpart inside 
the error circles (radius=1.2$\arcsec$) to our limiting magnitude of $26.2$ with the F606W.

SN 2001fu, M-03-23-011:  There are no HST images of this object.

SN 2002hl, NGC 3665:  No WFPC2 or ACS images of this object exist. There is a NICMOS image, however the SN does not lie within the image.

SN 2003gs, NGC 936:  Using the ACS images, we were able to align the position of the SN with WFPC2 images 
taken 7 years before the SN, and another ACS image taken 2 years after so that the error circle had radius $0.2\arcsec$. 
None of these show any object within the SN error circle to the limiting magnitude of $24.6$ in the F555W.

SN 2003hv, NGC 1201:  There are many ACS images of the supernova, but none of the region after the SN faded. 
Only one WFPC2 image was taken seven years before the SN. To find the SN region, we measured the offset of the SN from nearby point sources in the ACS image. 
We transferred this offset to the WFPC2 image. This yielded a very small error circle radius of $0.1\arcsec$.  At that location, we found a dim object with a S/N $\approx$ 3-4 and an apparent magnitude 
of 22.6 in the F814W (Vegamag system); the exposure was only 160 seconds.  Assuming $V-I=1.1$ and a distance modulus of 31.53, the object 
has an absolute magnitude of $M_I= -8.9$ or $M_V=-7.8$, which would be consistent with that of a typical globular cluster. See fig. 1

SN 2003hx, NGC 2076:  No HST images have been taken of this object.

SN 2003if, NGC 1302:  No HST images have been taken of this object.

SN 2004W, NGC 4649:  The ACS image fails to show an object within the error circle (radius = 1.0$\arcsec$) at our limiting magnitude of 25.3 in the F850LP. The SN position did not fall on any WFPC2 images.

SN 2005cf, M-01-39-03:  All HST images of the galaxy contain the SN location and show the SN before it faded. There are no images of the SN location after the SN faded.

SN 2005cn, NGC 5061:  No HST images overlap with the SN position.

SN 2005cz, NGC 4589:  In the ACS images, there were no objects within our error circle (radius = 0.9$\arcsec$) to a limiting magnitude of 25.1 in the F814W filter.

SN 2006dd, NGC 1316:  
There were no objects visible in our error circle (radius = 1.2$\arcsec$) at our limiting magnitude of 24.7 with the F814W. 
We note that the region is in an area of significant dust extinction, which might obscure a low optical luminosity cluster on the far side of the galaxy.

SN 2006dy, NGC 5587:  There is a small patch of slightly brighter pixels within the error circle (radius =1.3$\arcsec$), but this is not a significant detection. The limiting magnitude is 25.3 in the F814W.

SN 2006E, NGC 5338:  No HST images of this object exist.

SN 2007gi, NGC 4036:  WFPC2 images include this SN.  No optical counterpart was found within the error circle (radius = 0.9$\arcsec$) to our limiting magnitude of 21.8 using the F658N filter.

SN 2007on, NGC 1404:  This target was observed with ACS images and from this data set 
we determined an error circle of radius = 1.0$\arcsec$ and a limiting magnitude of 
25.0 with the F850LP filter.  We concur with the thorough treatment 
of \citet{voss08} that there is no optical counterpart at the location of the supernova.  
Initially, \citet{voss08} claimed that the supernova had an X-ray precursor, but with improved 
data, they determined that the faint X-ray source is not aligned with the SN Ia event \citep{roel08}.

\section{Discussion and Conclusions}

Our goal was to determine whether SN Ia events are produced through dynamical interactions in globular clusters, since globular clusters often host most of the close binary systems composed of a neutron star and a normal star.  To address this, we examined 36 galaxies that have had SN Ia events and are sufficiently close that most globular clusters can be detected with HST images. A number of galaxies proved to be unsuitable in that the SN Ia event fell outside of the image, or the images include the bright supernova, which prevented us from detecting an underlying globular cluster (there was no pre-event image or an image after the supernova had faded). 21 of the galaxies fulfilled all of our selection criteria and were fully considered.
Four of the objects had positional uncertainties too high to be of use, leading to multiple objects within the error circle.
Of the systems with adequate data and astrometry, we found 16 SN events without a definitive optical counterpart.  In only one event, SN 2003hv in NGC 1201, is there a possible optical counterpart found with low positional uncertainty.
The optical counterpart is not a high S/N detection, so the surface brightness profile cannot be determined nor can a color be calculated.  If the optical counterpart is at the distance of NGC 1201, it would have a magnitude of about $M_V=-7.7$, which is a typical globular cluster.
A deeper observation would show whether this is a reliable source and if it has the colors and extent expected from a globular cluster.

We can consider how these non-detections and a possible detection relate to the number of detections that might be theoretically predicted.  If globular clusters host SN Ia events at the same rate as the rest of the galaxy, we would not have expected any detections unless we used a sample size approximately 10 times larger.
We take the conservative assumption that all of the ambiguous or contamination cases are not true associations of a GC with a SN Ia. Based on this assumption, one detection in our sample of $21$ systems will imply a rate enhancement of about 10 for SN Ia events in GCs over field populations. This calculation was made by  dividing the total luminosity of the observed GCs by the total luminosities of their respective host galaxies. Total system GC luminosities were found by multiplying the number of GCs in each galaxy by the mean luminosity given by the Milky Way Globular Cluster Luminosity Function (see Table 7 for data on the GC systems and the data sources). When the number of globular clusters in a galaxy was not available in the literature, we estimated the number by counting the number of GCs in the systems using WFPC2 images. 

Not all globular clusters could be detected with the available images due to the depth of the
image, which is determined by a combination of the exposure time, filter used, and stellar background,
which rises toward the center of the galaxy. Keep in mind that this method of estimation is inherently biased, as we only have access to data on the GC light that falls on the HST field of view.
For each SN, we determined the local background in the image and calculated the limiting 
4$\sigma$ detection threshold.  Using this threshold, we calculated the accessible fraction of the 
stars given by the globular cluster luminosity function, using the luminosity function for the
Milky Way and M31, corrected to the relevant wavelength band \citep{binney1998, ashman1995}.
The typical object has observations that included more than 95\% of the globular cluster stars
with only a few exceptions in which the observation time is short (less than 200 seconds; the median
observation is 900 seconds) or the SN occurs close to the inner part of the galaxy where the
central surface brightness is large.  Due to a few objects where less than 60\% of the globular cluster
stars are detectable, 85\% of globular cluster stars would have been detected.  We include this
incompleteness correction in formulating the statistics below.

This approach yielded similar numbers for galaxies with known globular cluster data \citet{pfahl09}.
If none of the ambiguous cases have an optical counterpart, then there are 0 globular clusters 
at the locations of SNe Ia for 21 events.  This result implies a correspondence rate (Poisson 
mean) of $\mu$ = 0.0 - 0.18 between SNe Ia and globular clusters, at the 95\% confidence level (using the Wilson score interval, \citealt{wilson27}). 
If one ambiguous object has a globular cluster counterpart (SN 2003hv is the most likely), then the
implied correspondence rate is $\mu$ = 0.048, with a 95\% confidence bound of $\mu$=0.01-0.23.  
A value for $\mu$ of 0.18 (the upper bound to the confidence range for zero coincidences)
is a rate enhancement of $33$. A value for $\mu$ of 0.23 (the upper bound to the confidence range for one coincidence) is a rate enhancement of $42$. 

For the ambiguous cases that we consider less likely, two detections would result in a correspondence rate of $\mu$=.095 with a 95\% confidence bound of  $\mu$=.03-.29 which will result in a maximum rate enhancement of $53$. 
Three detections would result in a correspondence rate of $\mu$=.14 with a 95\% confidence bound of $\mu$=.05-.34 which will result in a maximum rate enhancement of $62$. Four detections would result in a correspondence rate of $\mu$=.14 with a 95\% confidence bound of $\mu$=.07-.40 which will result in a maximum rate enhancement of $73$. Five detections would result in a correspondence rate of $\mu$=.24 with a 95\% confidence bound of $\mu$=.10-.45 which will result in a maximum rate enhancement of $82$.

In a recent work, \citet{voss12} also examined the relationship between SNe Ia and globular clusters.  
They included events out to a distance of 100 Mpc and performed both a literature search 
as well as using archival HST data.  They did not restrict their study to early-type galaxies 
and most of the galaxies in their sample are spirals.  There were 35 SNe Ia examined, with 
18 systems close enough to identify most of the globular clusters.  For the remaining 17, 
only the brightest globular clusters would have been visible.  They do not find a 
globular cluster counterpart to any of the SNe Ia.  Unlike our study, no ambiguous cases 
are reported.  They present results as a function of galaxy type, so for the early-type 
galaxies with normal SN Ia events, the subsample corresponding to our sample, they 
have an effective 7.25 objects, and they place an upper limit to the enhancement rate 
of 46-90 (90\% confidence) and 82-163 (99\% confidence).  These results are consistent 
with ours, although we have a larger sample of SN Ia events in early-type galaxies.

Our limits on the rate enhancement is much less than that for luminous X-ray binaries ($> 10^{35} erg s^{-1}$), where the enhancement for globular clusters in early type galaxies is about 200 (see \S 1).  
If the rate enhancement of 200 also applied to SN Ia events, approximately 
one-third of these supernovae would be associated with globular clusters.
We can compare our results to those from theoretical modeling.
In their theoretical stellar evolution modeling of white dwarfs in binary systems, \citet{ivan06} 
find that globular clusters can produce as many SN Ia events as the rest of the galaxy.  
For the galaxy and globular cluster parameters chosen, this corresponds to an enhancement rate
of about 200. 
\citet{shara02} used N-body simulations and found that the double-degenerate channel for SN Ia formation would be enhanced in globular clusters by a factor of 10 or more.
\citet{pfahl09} discuss this issue, and based on a variety of calculations, they suggest an enhancement of 10 for the production of SN Ia events in globular clusters.
Our results are in conflict with a high enhancement rate, such as suggested by \citet{ivan06}, 
but are consistent within a rate enhancement of an order of magnitude. 

The rate at which SNe are associated with globular clusters may depend on radius within 
the galaxy, which can lead to a bias in our results.  The globular cluster distribution 
does not necessarily follow the light of the galaxy in the sense that there are relatively 
fewer globular clusters in the inner region, as is the case in the giant elliptical galaxy 
NGC 4486 (M 87), where the core radius for the distribution is 56$\arcsec$, compared to 
6$\arcsec$ for the underlying stellar component \citep{kundu99}.  If the relative deficit 
of globular clusters in the inner part of early-type galaxies were due to dynamical 
destruction (e.g., \citealt{hut92}), a potential SN Ia progenitor, originally created 
in a globular cluster, would appear as part of the underlying stellar distribution.  
Because the \textit{HST} observations exist primarily for the inner 1$^{\prime}$ of the 
galaxies in our sample, this could lead to an underestimation of the true association rate 
between globular clusters and SNe.

While this might seem like a plausible scenario, observational tests of globular cluster 
destruction do not provide support.  Destruction mechanisms should act at different 
rates for low and high mass globular clusters, which should modify the globular cluster 
luminosity function \citep{agui88, murray92}.  This would be manifest as a radial change 
in the luminosity function, but none is seen \citep{kundu99}.  
The other element needed to cause a bias is that globular clusters are destroyed 
after SN Ia progenitors are formed.  However, the typical time since destruction is 
several Gyr while the binary formation mechanism in globular clusters occurs 
continuously \citep{ivan06} and on the much shorter relaxation time of 10$^8$-10$^9$ yr. 
We conclude that there is not an obvious central galaxy bias for finding GC-SN 
associations, but a more definitive statement could be made by accumulating 
statistics on SN Ia events occurring more than 1$^{\prime}$ from the galaxy center, 
for which there is currently little \textit{HST} data.

If globular clusters had proven to be SN Ia factories, it would have solved a problem
with the production rate of these systems.  
In his review, \citet{maoz10} summarizes the arguments that for SN Ia in 
old populations, the double-degenerate channel is the likely path.
However, the present-day rate for the double-degenerate path in the field stars of 
early-type galaxies falls an order of magnitude short of the observed rate of SN Ia events.  
Now that we have shown that globular clusters are not responsible for most SN Ia events 
(barring unforeseen bias effects), the difficulty of forming enough of these events persists.

If the enhancement rate for SNe Ia in globular clusters is about 10, it should be 
possible to improve the size of the sample by at least a factor of two and 
confirm this prediction.  
Within our sample, a dozen targets have no image data, and with current instruments 
on HST, the needed imaging can be accomplished with only modest time investments.  
Other targets will become available in the near future as the SN Ia events fade.  
Perhaps the most urgent observation is to resolve the tentative association of a 
supernova with an optical counterpart in NGC 1201.  As this has only a single 160 
second exposure, substantially improved images can be obtained easily and 
two-color images would determine the color and morphology of the object.
This is a program worth pursuing, because identifying even a single SN Ia with a 
globular cluster gives us the age and metallicity of the progenitor \citep{pfahl09}, 
information that is otherwise unobtainable.  Such information not only gives insight 
into the formation mechanism of SNe Ia for old populations, it might help reduce 
systematics when using SNe Ia for cosmology.

\section{Acknowledgements}
The authors would like to thank Mario Hamuy, Maximilian Stritzinger, Carlos Contreras, 
Ted Dobosz, Stefano Benetti, Mark Philips, and Chris Smith for providing us with 
unpublished images and information on the supernovae used in this work.  Also, we 
would like to thank Jimmy Irwin, Lou Strolger, and Roger Chevalier for their comments and insight.  
This work is based on observations made with the NASA/ESA Hubble Space Telescope, 
obtained from the data archive at the Space Telescope Science Institute. STScI is 
operated by the Association of Universities for Research in Astronomy, Inc. under 
NASA contract NAS 5-26555.
Also, this research has made use of the NASA/IPAC Extragalactic Database (NED) 
which is operated by the Jet Propulsion Laboratory, California Institute of Technology, 
under contract with the National Aeronautics and Space Administration.
This work has made use of SAOImage DS9, developed by Smithsonian Astrophysical Observatory.
Finally, we would like to acknowledge financial support from NASA for these activities through a ADP grant.
We gratefully acknowledge financial support for this project from NASA through an ADAP grant.


\begin{deluxetable}{rrrrrrr}
\tablecolumns{7}
\tablewidth{0pc}
\tablecaption{Nearby Supernova and Galaxy Association of Events with Sufficient Data}
\tablehead{
\colhead{SN event} & \colhead{Galaxy} & \colhead{SN Ra}& \colhead{SN Dec} & \colhead{Offsets}&  & \colhead{Galaxy Type}}
\startdata
1961    H&      NGC4564& 12h36m27.265s &11d26m24.62s&0&5N&      Ea \\
1976    K&      NGC3226&10h23m28.826s &19d54m15.14s&28E&21N&       E2: pec\\
1983    O&      NGC4220&  12h16m13.204s&47d52m45.16s&19E& 15S&     SA0$^{+}$(r)\\
1986    G&      NGC5128& 13h25m36.458s&-43d01m53.65s&120E&60S&       S0 pec\\
1991    bg&     NGC4374&  12h25m03.698s&12d52m15.64s&2E&57S&      E1\\
1991    F&      NGC3458&  10h56m15.099s &57d06m54.34s&13.4E&12.1S&    SAB0\\
1992    A&      NGC1380&03h36m27.414s  &-34d57m31.39s&3W&62N&      SA0\\
1994    D&      NGC4526&12h34m02.395s &07d42m05.70s&9W&7N &     SAB0$^{0}$(s)\\
1996    X&      NGC5061&13h18m01.130s &-26d50m45.30s&52W&31S&       E0\\
1996    bk&     NGC5308& 13h46m57.980s&60d58m12.90s&17.9W&10.5S&       S0\\
1999    gh&     NGC2986& 09h44m19.750s &-21d16m25.00s&52W&15.8N&      E2\\
2000    ds&     NGC2768&  09h11m36.240s &60d01m42.20s&9.1W&32.1S&     E6\\
2001    A&      NGC4261&12h19m23.010s&05d49m40.50s&3.1W&10.7N&        E2-3\\
2003    gs&     NGC 936&02h27m38.360s  &-01d09m35.40s&13.4E&14.6S&      SB0$^{+}$(rs)\\
2003    hv&     NGC1201& 03h04m09.320s &-26d05m7.5s&17.2E&56.7S&      SA0$^{0}$(r)\\
2004    W&      NGC4649&12h43m36.520s   &11d31m50.80s&51.6W&78.7S&     E2\\
2005    cz&     NGC4589& 12h37m27.850s&74d11m24.50s&12E&6S&       E2\\
2006    dd&     NGC1316&03h22m41.620s   &-37d12m13.00s&0.3W&16N&     SAB0$^{0}$(s) pec\\
2006    dy&     NGC5587& 14h22m11.450s &13d55m14.20s&10E&9N&      S0/a\\
2007    gi&     NGC4036 &12h01m23.420s &61d53m33.80s  &23E&11S&    S0$^{-}$\\
2007    on&     NGC1404&03h38m50.900s & -35d34m30.00s&12E&68N&      E1\\
\enddata
\end{deluxetable}

\begin{deluxetable}{rrrrrrr}
\tablecolumns{7}
\tablewidth{0pc}
\tablecaption{Nearby Supernova and Galaxy Association of Events with Insufficient Data}
\tablehead{
\colhead{SN event} & \colhead{Galaxy} & \colhead{SN Ra}& \colhead{SN Dec} & \colhead{Offsets}&  & \colhead{Galaxy Type}}
\startdata
1919    A&      NGC4486 (M87)& 12h30m48.64s  &12d25m04.2s&15W & 100N &  cD-1 pec\\
1960    R&      NGC4382&12h25m24.834s &18d09m19.35s&8E&132S&       SA0$^{+}$(s) pec\\
1980    I&      NGC4374& 12h25m38.117s &12d53m23.50s&454E&20N&      E1\\
1982    W&      NGC5485&  14h07m08.958s&55d01m07.38s&19W&63N&      SA0 pec\\
1983    G&      NGC4753& 12h52m20.787s  &00d39m36.72s&17W&14S&    I0\\
1995    D&      NGC2962& 09h40m54.753s  &05d08m26.16s&11E& 90.5S &   (R)SAB0$^{+}$(rs)\\
1997    X&      NGC4691& 12h48m14.280s  &-03d19m58.50s&7.2E&0.3N&     (R)SB0/a(s) pec\\
2000    cx&     NGC 524& 01h24m46.150s &09d30m30.90s&23.0W&109S&      SA0$^{+}$(rs)\\
2001    fu&     M-03-23-11&  08h52m16.580s&-17d44m29.80s&25.1W&10.6N&   (R')SB0$^{0}$(s)\\
2002    hl&     NGC3665& 11h24m40.120s &38d46m03.00s&42.0W&16.6N&      SA0$^{0}$(s)\\
2003    hx&     NGC2076& 05h46m46.970s&-16d47m00.60s&5.2W&2.6S&       S0$^{+}$ \\
2003    if&     NGC1302&03h19m52.610s &-26d03m50.50s&19.3E&11.9S&       (R)SB0/a(r)\\
2005    cf&     M-01-39-03&15h21m32.210s &-07d24m47.50s &15.7W&123N&    S0 pec\\
2005    cn&     NGC5061&13h18m00.460s &-26d48h33.10s&62W&103N&      E0\\
2006    E&      NGC5338 & 13h53m28.650s&05d12m22.80s &31.1E&6.3S&      SB0\\
\enddata
\end{deluxetable}

\begin{deluxetable}{rrrrrrr}
\tablecolumns{7}
\tablewidth{0pc}
\tablecaption{Important Images, Exposure Times, and Filters}
\tablehead{
\colhead{SN event} & \colhead{Galaxy} & \colhead{Instrument} &\colhead{Detector}& \colhead{Image} & \colhead{Exp. Time (s)} & \colhead{Filter}}
\startdata
1961    H&      NGC4564&ACS& WFC &jfs22011&1120&F850LP \\
1976    K&      NGC3226&ACS& WFC &j6jt02y2q&350&F814W\\
1983    O&      NGC4220& ACS & WFC &j8mx74010&700&F658N\\
1986    G&      NGC5128& WFPC2 &1 &u4100206b&753&F555W\\
1991    bg&     NGC4374&  WFPC2 & 1&u34k0103t&4320&F814W/F658N/F547M\\
1991    F&      NGC3458&  WFPC2 & 1&u67n3202b&160&F606W \\
1992    A&      NGC1380& WFPC2 &1 &u3gh0702b &2600&F555W\\
1994    D&      NGC4526&ACS & WFC &jfs08011&1120&F850LP\\
1996    X&      NGC5061&STIS& MIRVIS &o6fz12010&900&MIRVIS\\
1996    bk&     NGC5308& ACS& WFC&j9ew04030&3480&F814W\\
1999    gh&     NGC2986& WFPC2&1 &u3cm1402r&700&F702W \\
2000    ds&     NGC2768&  ACS & WFC&j6jt08011&1700&F658N\\
2001    A&      NGC4261& ACS &HRC&j8da04031 &782&F606W\\
2003    gs&     NGC 936& WFPC2&1&u2jf0303t&140&F555W\\
2003    hv&     NGC1201& WFPC2 &1& u2tv 3501b&160&F814W\\
2004    W&      NGC4649& ACS & WFC& j8fs03011&1120&F850LP\\
2005    cz&     NGC4589& ACS & WFC&j9ar03030&1600&F814W\\
2006    dd&     NGC1316& ACS & WFC&j6n201030&2200&F814W\\
2006    dy&     NGC5587& WFPC2 &1&u41v0704m&890&F814W/F450W\\
2007    gi&     NGC4036 & WFPC2 &1&u3780303b&700&F658N\\
2007    on&     NGC1404&ACS & WFC&j90x3010&1130&F850LP\\
\enddata
\end{deluxetable}

\begin{deluxetable}{rrrr}
\tablecolumns{4}
\tablewidth{0pc}
\tablecaption{Region Errors and Limiting Magnitudes of Useable Events}
\tablehead{
\colhead{SN event} & \colhead{Galaxy} & \colhead{SN Region Error (arcsec)} &\colhead{Limiting Magnitude (mag)}}
\startdata
1961    H&      NGC4564&1.7& 23.0 \\
1976    K&      NGC3226&.1& 25.1\\
1983    O&      NGC4220& 1.8& 22.3 \\
1986    G&      NGC5128& 1.7 & 23.2 \\
1991    bg&     NGC4374&  1.7 & 24.4\\
1991    F&      NGC3458&  1.0 & 24.4\\
1992    A&      NGC1380& 2.5 &24.0 \\
1994    D&      NGC4526&.4 & 23.2 \\
1996    X&      NGC5061&.2& 26.0 \\
1996    bk&     NGC5308& 2.0& 25.3\\
1999    gh&     NGC2986& .9& 25.3\\
2000    ds&     NGC2768&  2.0 & 24.2\\
2001    A&      NGC4261& 1.2 &26.2\\
2003    gs&     NGC 936& .2&24.6\\
2003    hv&     NGC1201& .1 &22.6\\
2004    W&      NGC4649& 1.0& 25.3\\
2005    cz&     NGC4589& .9 & 25.1\\
2006    dd&     NGC1316& 1.2 & 24.7\\
2006    dy&     NGC5587& 1.3 &25.3\\
2007    gi&     NGC4036 & .9 &21.8\\
2007    on&     NGC1404& 1.0 & 25.0\\
\enddata
\end{deluxetable}

\begin{deluxetable}{rrrr}
\tablecolumns{3}
\tablewidth{0pc}
\tablecaption{Archival Image Materials: ACS Observations}
\tablehead{
\colhead{SN event} & \colhead{Galaxy} & \colhead{ACS Data Set(s) Used} & \colhead{Observation Dates} }
\startdata
1961    H&      NGC4564&        J8FS22ETQ       &7/7/03\\
1976    K&      NGC3226 &J6JT02021,J6JT02011&   3/8/2003, 3/8/2003\\
1980    I&      NGC4374 &J8FS06O0Q, J8FS06011, J8FS06021        &1/21/03, 1/21/03,1/21/03\\
1982    W&      NGC5485 &J8MXC5010, J8MXC5FDQ   &11/8/03, 11/8/03\\
1986    G&      NGC5128&        None    &N/A\\
1991    F&      NGC3458&        None    &N/A\\
1991    bg&     NGC4374 &J8FS06O0Q, J8FS06011, J8FS06021        &1/21/03, 1/21/03,1/21/03\\
1992    A&      NGC1380 &J90X04010, J90X04020, J90X04DJQ        &9/6/04, 9/7/04, 9/6/04\\
1994    D&      NGC4526 &J8FS08LNQ&     7/12/03\\
1996    X&      NGC5061 &O6FZ12010 (STIS image, no ACS available)&      5/5/02\\
1999    gh&     NGC2986 &None&  N/A\\
2000    ds&     NGC2768 &J6JT08011, J6JT08021, J8DT02011, &     1/14/03, 1/14/03, 5/31/02,\\
2001    A&      NGC4261 &J9OB01010, J9OB01020&  12/25/06, 12/25/06\\
2003    gs&     NGC 936 &J8Z457011&     9/29/04\\
2003    hv&     NGC1201&        J8Z465021&      7/12/04\\
2004    W&      NGC4649 &J8FS03011, J8FS03021&  6/17/03, 6/17/03\\
2005    cz&     NGC4589 &J9AR03011, J9AR03021, J9AR03031&       11/11/06, 11/1106, 11/11/06\\
2006    dd&     NGC1316 &J6N202010, J6N201010, J6N201030&       3/4/03, 3/7/03, 3/7/03 \\
2006    dy&     NGC5587 &None&  N/A\\
2007    gi&     NGC4036 &None&  N/A\\
2007    on&     NGC1404&        J90X03010, J90X03020, J90X03AXQ &       9/10/04, 9/10/04, 9/10/4\\
\enddata
\end{deluxetable}

\begin{deluxetable}{rrrr}
\tablecolumns{3}
\tablewidth{0pc}
\tablecaption{Archival Image Materials: WFPC2 Observations}
\tablehead{
\colhead{SN event} & \colhead{Galaxy} & \colhead{WFPC2 Data Set(s) Used} & \colhead{Observation Dates} }
\startdata
1961    H&      NGC4564 &U3HY0301B&     6/18/99\\
1976    K&      NGC3226 &None   &N/A\\
1980    I&      NGC4374&        U27L4P01B, U34K0103B, U34K0106B &4/4/1994, 3/4/96, 4/3/96\\
1982    W&      NGC5485 &None   &N/A\\
1986    G&      NGC5128 &U4100201B, U4100206B, U4100209B&       10/1/98,10/1/98,10/1/98\\
1991    F&      NGC3458 &U67N3202B&     6/24/01\\
1991    bg&     NGC4374 &U27L4P01B, U34K0103B, U34K0106B        &4/4/1994, 3/4/96, 4/3/96\\
1992    A&      NGC1380&        U3GH0702B       &9/4/96\\
1994    D&      NGC4526 &U3HY0801B&     6/18/99\\
1996    X&      NGC5061&        U3GH0801B, U3GH0803B, U3GH0805B&        9/1/97, 9/1/97, 9/1/97\\
1999    gh&     NGC2986 &U3CM1402B      &1/18/99\\
2000    ds&     NGC2768 &U3M71605B, U3M71608B, U2TV1801B&       20/5/98, 20/5/98, 4/12/95\\
2001    A&      NGC4261 &U2I50205B      &13/12/94 \\
2003    gs&     NGC 936 &None   &N/A\\
2003    hv&     NGC1201&        U2TV3501B&      27/12/96 \\
2004    W&      NGC4649 &None&  N/A\\
2005    cz&     NGC4589 &U2BM0B03B, U3CM7302B, U67G7701B        &14/05/94, 1/10/99, 6/23/01\\
2006    dd&     NGC1316 &None   &N/A\\
2006    dy&     NGC5587 &U41V0702B      &13/7/97\\
2007    gi&     NGC4036 &U3780303B      &12/5/97\\
2007    on&     NGC1404 &U34M0408B, U34M0402B, U29R5J01B&       4/3/96, 4/3/96, 12/06/95\\

\enddata
\end{deluxetable}

\begin{deluxetable}{rrrr}
\tablecolumns{3}
\tablewidth{0pc}
\tablecaption{Globular Cluster Populations in Galaxies}
\tablehead{
\colhead{SN event} & \colhead{Galaxy} & \colhead{Number of GCs} & \colhead{Reference} }
\startdata
1961    H&      NGC4564 & 2700& \citet{kiss97} \\
1976    K&      NGC3226 & 480 & \citet{kiss97} \\
1986    G&      NGC5128 & 1500& \citet{har98} \\
1991    bg&     NGC4374& 3040& \citet{kiss97} \\
1992    A&      NGC1380& 2300& \citet{kiss971} \\
1996    X&      NGC5061& 84& \citet{geb99} \\
2000    ds&     NGC2768 & 118& \citet{kun01b} \\
2001    A&      NGC4261 & 321& \citet{gio05} \\
2003    hv&     NGC1201& 77& \citet{kun01b} \\
2004    W&      NGC4649 & 5100&  \citet{kiss97} \\
2005    cz&     NGC4589 & 179 & \citet{kun01a} \\
2006    dd&     NGC1316 & 1496& \citet{goo04} \\
2007    on&     NGC1404 & 138& \citet{geb99} \\

\enddata
\end{deluxetable}

\clearpage


\begin{figure}

\plotone{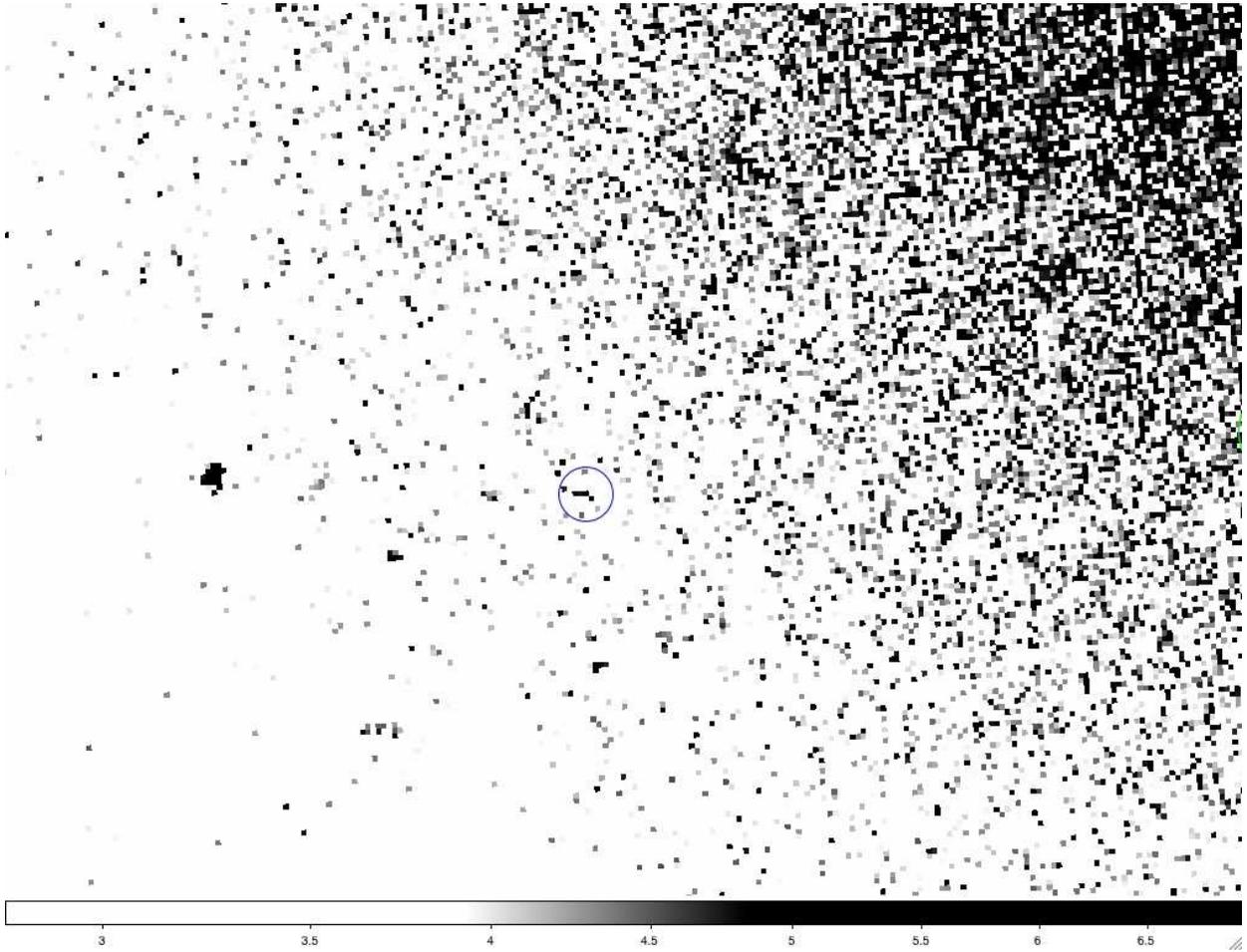}
\caption{The positional uncertainty of the Type Ia SN2003hv is approximately 0.1$\arcsec$. In this image, the SN Ia location is given by the circle, which has been enlarged from the true error circle size to show the object within it more clearly. The possible source within the error circle is at the 3-4$\sigma$ level.  We regard this as a possible but ambiguous counterpart in our analysis. The image is in WCS coordinates (north is up) and covers an $7.8\arcsec \times 5.2 \arcsec$ region.}
\end{figure}

\begin{figure}
\plotone{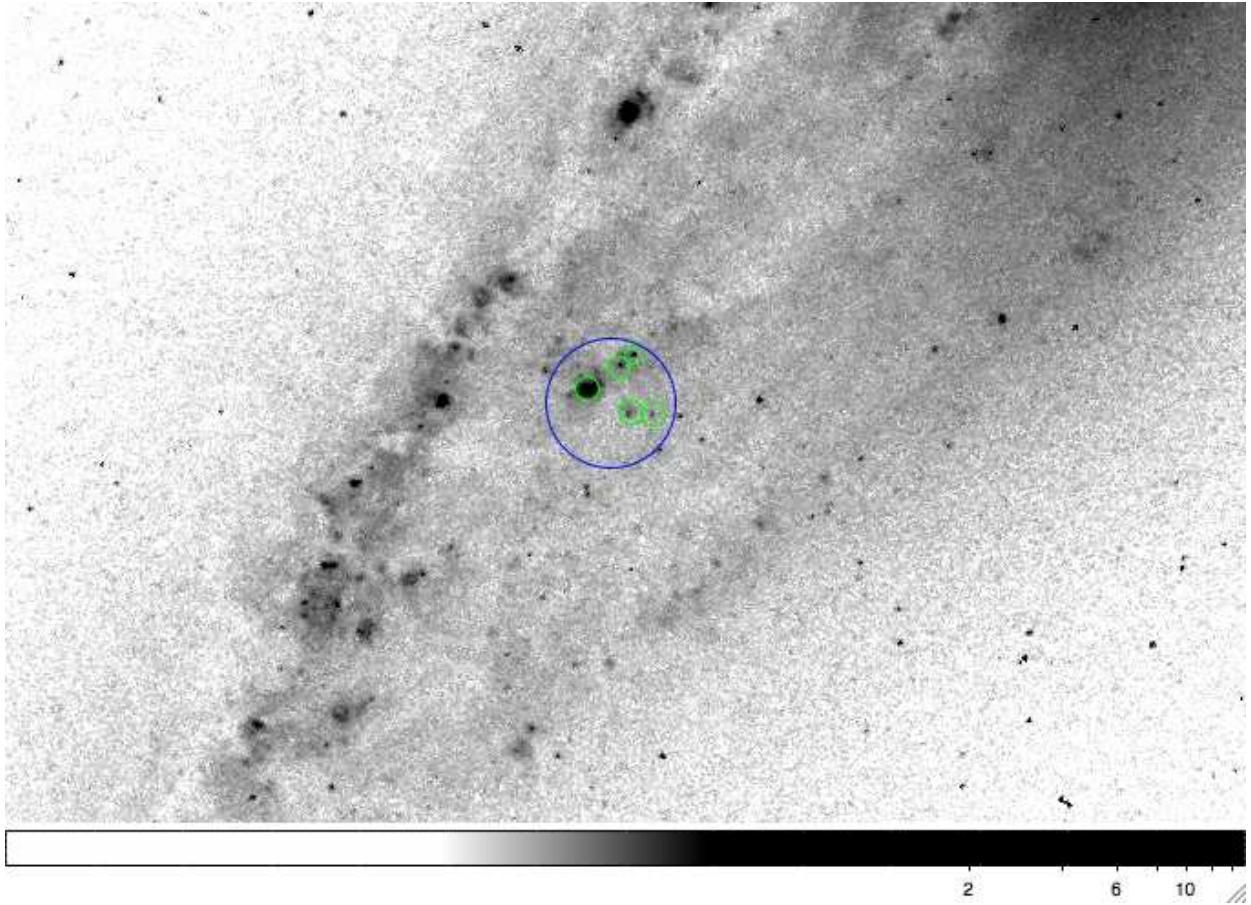} 
\caption{The location of SN1983O in NGC 4220 is given by the blue error circle, which has a radius of 1.8$\arcsec$. There are five optical sources marked within the error circle. The image size is $34\arcsec \times 23 \arcsec$ region.}
\end{figure}


\begin{figure}
\plotone{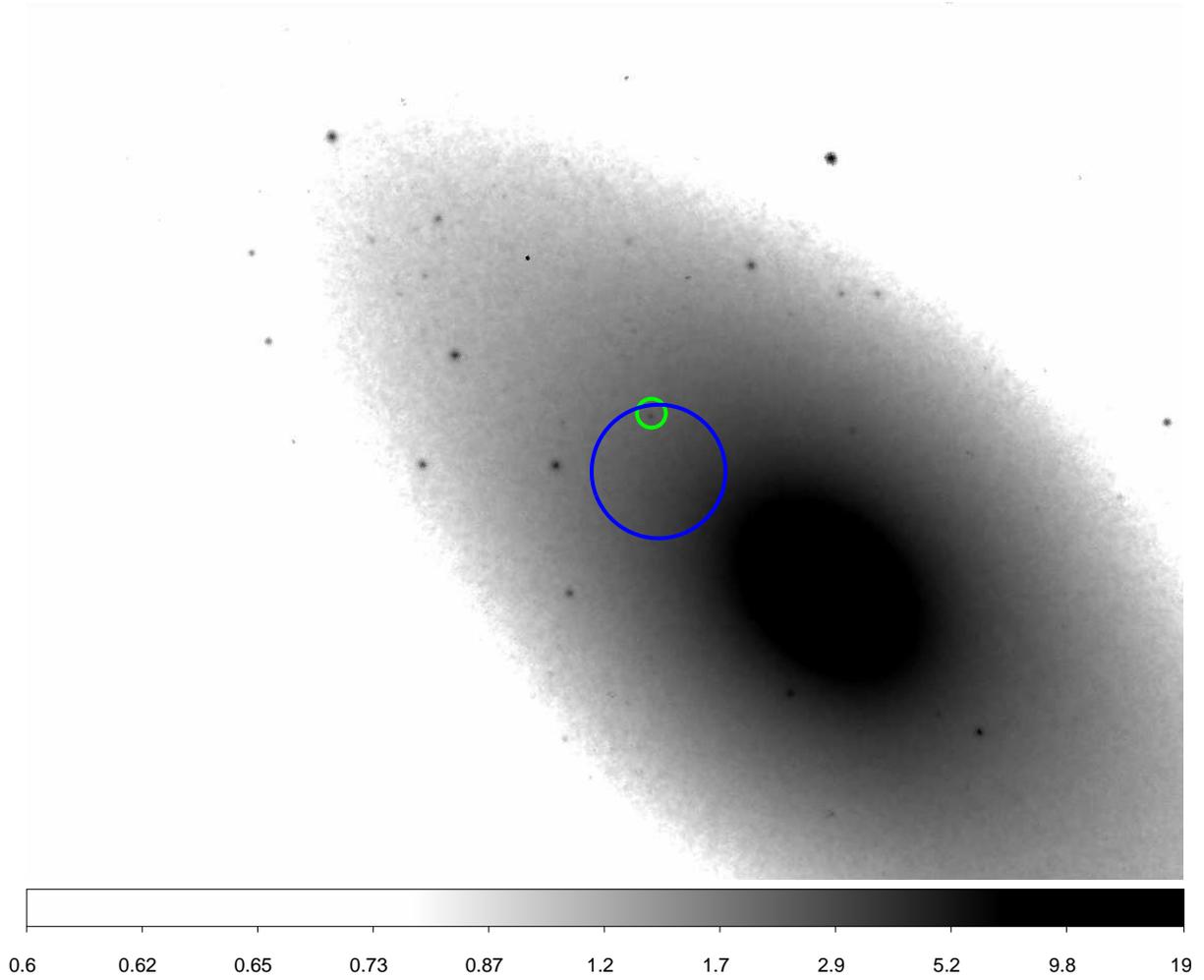} 
\caption{The location of SN1961H, which lies only 6$\arcsec$ from the center of NGC 4564, is given by the blue error circle, with a radius of 1.7$\arcsec$. There is one source marked within the error circle but this is likely a chance coincidence, given the background density near the center and the large error circle. The image is $29\arcsec \times 23 \arcsec$.}
\end{figure}

\begin{figure}
\plotone{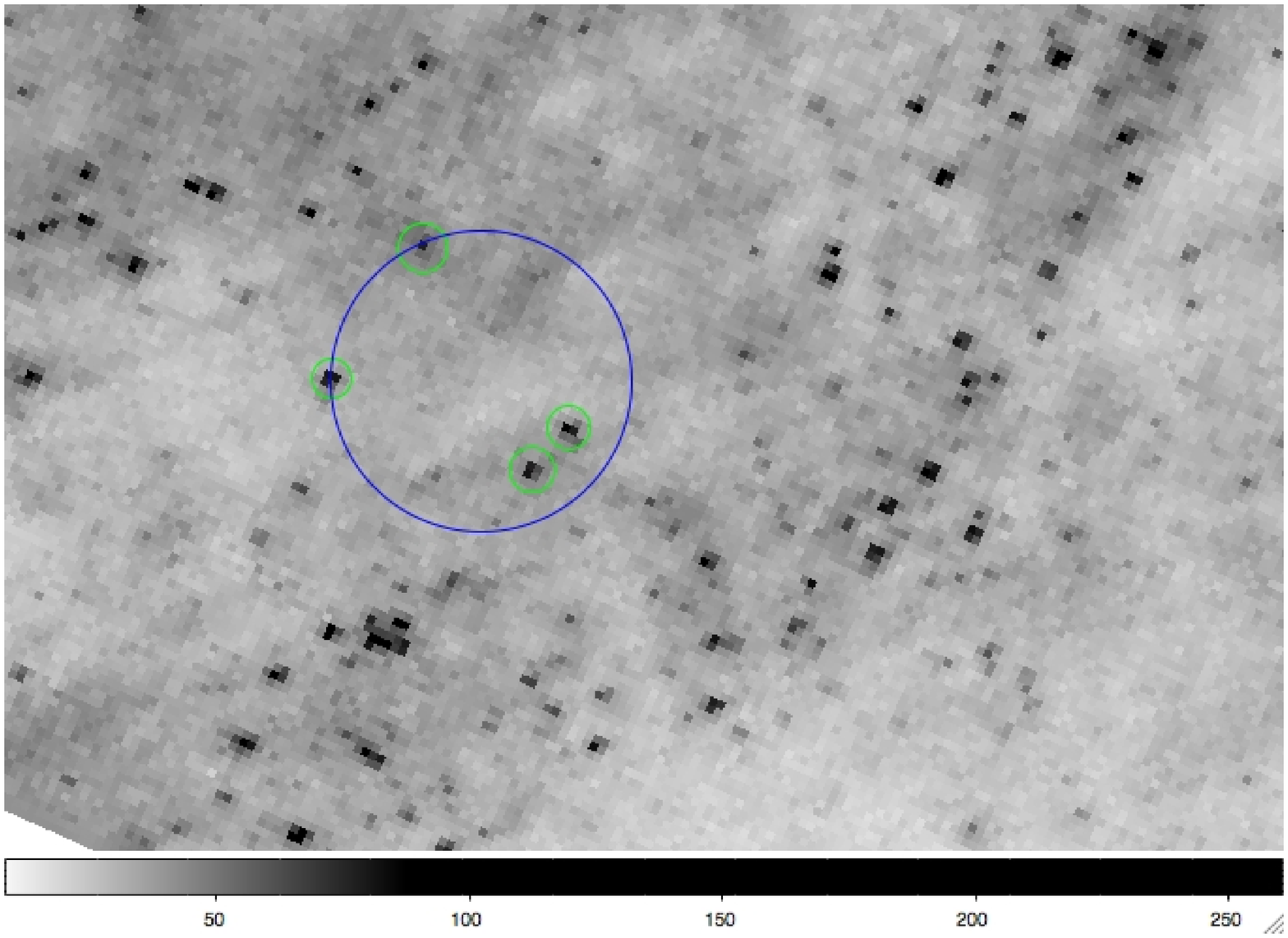} 
\caption{The location of SN1986G in NGC 5128 is given by the blue error circle, which has a radius of 1.7$\arcsec$. The source density is high and there are several objects marked within the error circle. The image is $16\arcsec \times 10\arcsec$.}
\end{figure}

\begin{figure}
\plotone{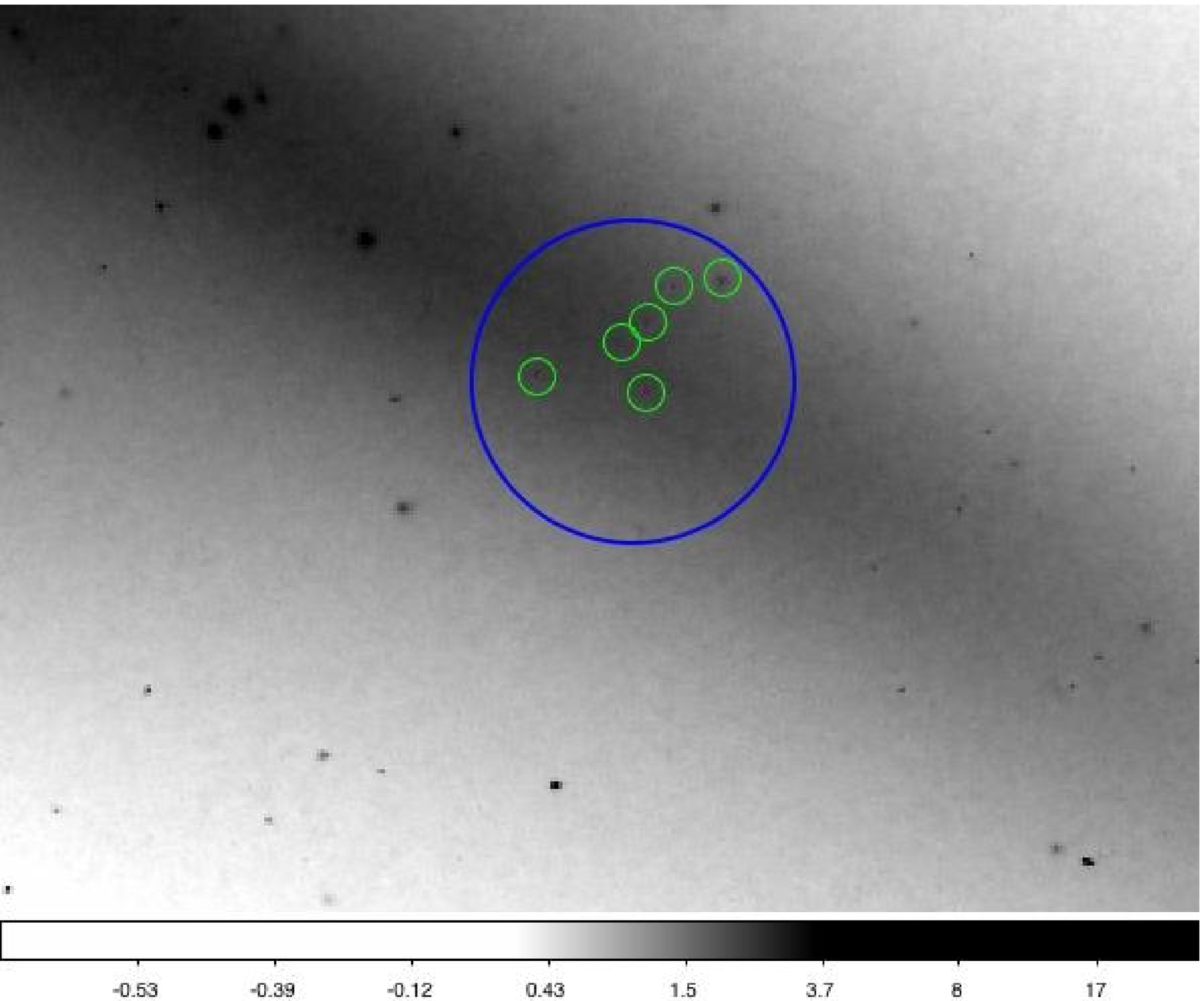} 
\caption{The location of SN1996bk in NGC 5308 is given by the blue error circle, which has a radius of 2.0$\arcsec$. The source density is high and there are several objects marked within the error circle. The image is $16\arcsec \times 10 \arcsec$.}
\end{figure}

\end{document}